# Ultrafast scintillating metal-organic frameworks films.

Lorena Dhamo[1], Jacopo Perego[1,†], Irene Villa[1,†], Charl X. Bezuidenhout[1,†], Ilaria Mattei[2], Alessia Landella[1], Silvia Bracco[1], Angiolina Comotti[1]* & Angelo Monguzzi[1]*

[1]Dipartimento di Scienza dei Materiali, Università degli Studi Milano-Bicocca, via R. Cozzi 55, Milano 20125, Italy

[2] INFN, Sezione di Milano, via G. Celoria 16, Milano 20133, Italy

**Abstract**.

Compositionally engineered metal-organic frameworks (MOFs) have been designed and used to fabricate ultrafast scintillating films with emission in both the UV and visible regions. The inclusion of hafnium (Hf) ions in the nodes of the MOF increases the interaction cross-section with ionizing radiation, partially compensating for the low density of the porous material and dramatically increasing the system scintillation yield. The high diffusivity of bimolecular excitons within the framed conjugated ligands allows bimolecular annihilation processes between excited states that partially quench the MOF luminescence, resulting in ultrafast scintillation pulses under X-ray excitation with kinetics in the hundreds of picoseconds time scale. Despite the quenching, the gain in scintillation yield achieved by incorporating Hf ions is large enough to maintain the light yield of the films above $10^4$ ph/MeV under soft X-rays. These unprecedented high efficiencies and simultaneous ultrafast emission kinetics obtained at room temperature in a technologically attractive solid-state configuration, together with the versatility of its composition allowing for further application-specific modifications, place the MOF platform in a prominent position for the realization of the next generation of ultrafast scintillation counters for high-energy physics studies and medical imaging applications.

Scintillation counters consist of a scintillating material, which is able to produce a light pulse upon interaction with ionizing radiations or particles, coupled to a photodetector, able to transduce the optical signal in an electrical output count. Ultrafast efficient scintillators are now actively searched for several modern advanced applications where the detection of ionizing radiation and particles must be performed in an extremely short time with a good signal-to-noise ratio.[1] This is the case of forefront high-energy physics experiments, where extremely high event rates are generated and must be recorded avoiding pile up.[2] Additionally, faster detectors are also desirable when high quality images at low radiation flux are acquired for medical applications such as the ToF-PET cancer imaging technique, which requests tens of picoseconds of time resolution in the detection of the γ-rays emitted by patients' injected radiotracers.[3,4]

The realization of luminous ultrafast scintillators with tens of picosecond reaction speed is therefore strongly desired, but it is a highly demanding challenge. Indeed, many parameters dominate the yield and kinetics of the scintillation mechanism, and they are usually anti-correlated. The material chemical composition, structure, intra- and intermolecular interactions, crystal symmetry, atomic number Z, electronic and charge transport properties, emission yield and light transport characteristics as well as the device geometry and size, all of them affect the complex photophysics beyond the scintillation mechanism and ultimately set the intensity and speed of the emitted scintillation light pulse. Currently, materials that show proper control on all the required parameters do not exhibit good light output and ultrafast emission kinetics at room temperature. Neither bulk and nanostructured crystalline inorganic scintillators, nor organic chromophores and plastics have the required characteristics at once. Materials with fast prompt luminescence, such as Cherenkov emission, would be promising if not for their low light yield (LY, the number of photons produced for each MeV of deposited energy), preventing them from achieving reasonable output intensity. Semiconductor nanocrystals and other high atomic-number nanomaterials are now considered, especially when embedded into polymeric matrices.[5-10] However, their implementation is tough owing to the low stopping power of polymeric composites, dramatic emission self-absorption and the need for peculiar surface chemistry to optimize emission and chemical stability in the matrix. Conjugated systems can have a reasonably fast scintillation kinetics in the nanosecond time scale,[11-13] but inadequate stopping power for high-energy photons due to their low density.[14,15]

Engineered luminescent metal-organic frameworks (MOFs), nanoporous hybrid materials consisting of a crystalline framework of conjugated fluorescent ligands linked to metal-oxide cluster nodes (Fig.1),[16-24] exhibit fast and efficient luminescence properties. Specifically, these hybrid materials have been recently proposed as scintillators,[25-30] since they contain both high-Z elements and highly emissive conjugated ligands. Indeed, we demonstrated the unique radiosensitization effect of the MOF architecture, which enhances the scintillation yield by enabling a better harvesting and down-conversion into radiative states of the deposited energy.[31] The use of MOFs has several intrinsic advantages: the MOFs' emission spectrum can be tailored and tuned by ultrafast energy transfer processes through the ligand framework to obtain a huge Stokes shift, avoiding reabsorption and maximizing light output;[32-35] the composition of MOF linking nodes can be ameliorated to include heavier elements to enhance primary interactions with γ-rays and high-energy secondary electrons. Moreover, the controlled growth of MOF thin films[36-39] recently allowed the fabrication of sensors based on different transduction mechanisms, ultra-selective membranes and photo- and electro-catalysts.[36,40-42] Thus, MOF thin films hold great promise for realizing devices appealing for technological transfer.

In this work, we obtained solid-state films of ultrafast-emitting MOFs, UV-emitting or with large Stokes shift blue emission, based on hafnium (Hf) oxide clusters as linking nodes (Hf-MOF). As conjugated scintillating ligands we use 2',5'-dimethyl-[1,1':4',1''-terphenyl]-4,4''-dicarboxylate (TP) TP, as energy donor or direct UV emitter in homo-ligand MOFs, and 4,4'-(anthracene-9,10-diyl)dibenzoate (DPA) as energy acceptor and blue emitter in hetero-ligand MOFs (Fig.1a). These MOF films feature an unprecedented sub-nanosecond scintillation kinetics down to 150 ps and LY > $10^4$ ph MeV$^{-1}$ under soft X-rays at room temperature, thus surpassing most state-of-the-art and commercially available fast scintillators. The photophysical investigations demonstrate that the crystalline MOFs play a pivotal role in preserving the individual ligand electronic and emission properties in a controlled arrangement and at short distances (<1 nm), while simultaneously allowing ultrafast, diffusion-mediated bimolecular processes, such as energy transfer and singlet-singlet annihilation (SSA) between molecular excitons, that set their final characteristics as scintillators (Fig.1b,c). The results obtained demonstrate that the MOF platform is an ideal candidate to reach a breakthrough in the most advanced scintillation detectors technology and applications.

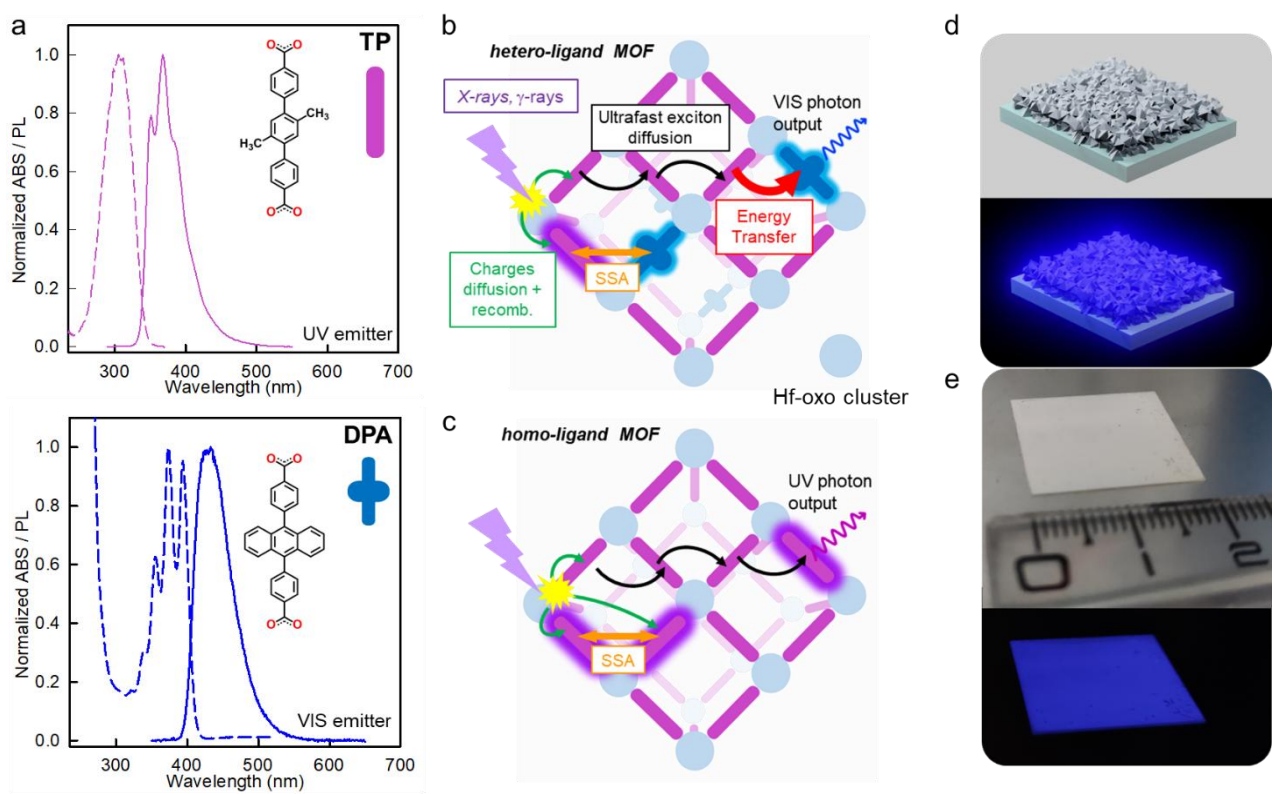

**Figure 1 | a,** Normalized absorption (ABS, dashed lines) and photoluminescence (PL, solid lines) spectra of the conjugated ligands employed to assemble **b,** homo-ligand and **c,** hetero-ligand scintillating MOFs based on Hf-oxo clusters as linking nodes. Arrows indicate the ultrafast processes occurring in the framework of conjugated ligands during the scintillation mechanism, including singlet-singlet annihilation (SSA, double headed arrow) between excited ligands in the singlet state. **d**, Sketch of MOF films grown on a glass substrate under ambient lighting and UV light. **e**, Digital image of a 2×2 cm$^2$ glass substrate coated with a ca 22 µm MOF film under ambient lighting and UV light.

**Preparation and structural characterization of Hf-MOF powders and films**

Highly luminescent Hf-MOF powders were prepared by a solvothermal process.[21,43] The homo-ligand MOF (denoted Hf-TP) was prepared using HfCl$_4$, benzoic acid (BA) as modulator, along with the linear, ditopic conjugated ligand, 2',5'-dimethyl-[1,1':4',1''-terphenyl]-4,4''-dicarboxylate (TP), as a direct UV emitter. The blue-emissive hetero-ligand MOF was obtained by the co-assembly of TP ligand with a low percentage of 4,4'-(anthracene-9,10-dial)dibenzoate (DPA), promoting effective energy transfer and increasing the Stokes shift.[44] The MOFs were activated by thermal treatment at 110 °C under high vacuum and their composition was determined from $^1$H solution NMR of the digested samples. In the hetero-ligand MOF a 1.6% molar fraction of DPA ligand was estimated (denoted Hf-DPA:TP-1.6%), and for both samples, the presence of BA was quantified to be 7-10% (Figs. S1-S3, Supplementary Tables 1-4). Infrared spectra of the MOFs displayed a peak at 1600 cm$^{-1}$ and a strong signal at 1412 cm$^{-1}$ associated with the asymmetric and symmetric stretching of the carboxylate coordinated to the Hf$^{4+}$ ions (Fig. S4). The crystal structures were elucidated from powder X-ray diffraction data using DFT methods and Rietveld refinement (Methods, Figs. S5, S6, Supplementary Table 5). The Hf-MOFs crystallize in the cubic space group *Fm*-3*m* and show short center-to-center ($d_{c-c}$) and phenyl-to-phenyl ($d_{Ph-Ph}$) distances between the ligands of 11.6 Å and 7.3 Å, respectively, which promote efficient energy diffusion and transfer within the framework (Fig. 2a, b). $^{13}$C CP MAS NMR spectra exhibited sharp resonances and a multiplicity of TP signals in agreement with the high symmetry of the unit cell (Fig.

2c, Figs. S7-S9). MOF crystals display well-defined octahedral morphology with a homogeneous particle size of 7.1±0.8 μm for Hf-TP and 6.5±1.2 μm for Hf-DPA:TP-1.6% (Figs. S10-S12). The MOF samples are highly stable up to 400 °C, as established by thermal gravimetric analysis, and the experimental residues are consistent with their compositions determined by NMR (Fig.S13).

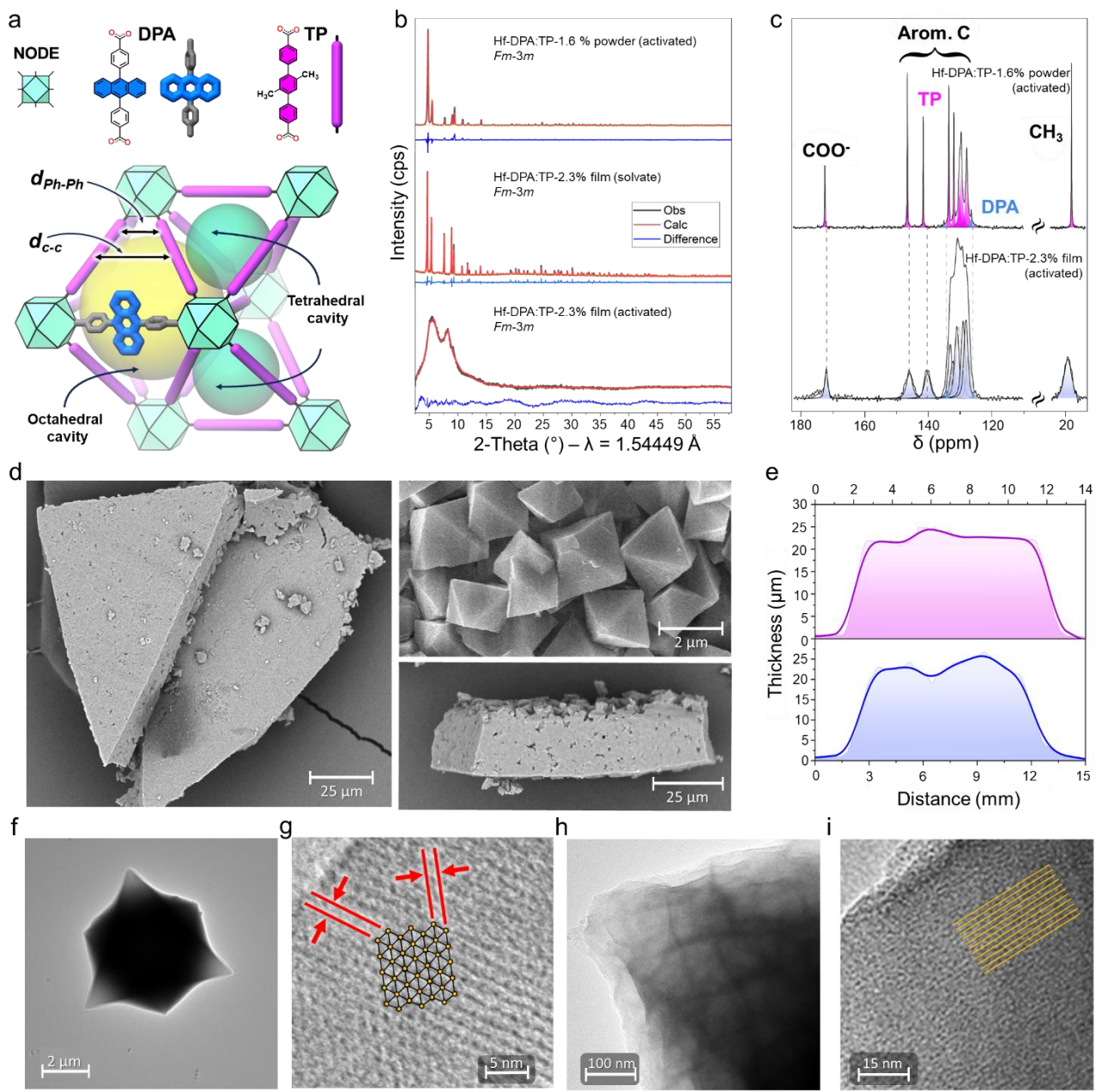

**Figure 2** | **a**, Hf-MOF composition and crystal structure highlighting the tetrahedral and octahedral cavities (green and yellow spheres, respectively) and the center-to-center ($d_{c-c}$) and phenyl-to-phenyl ($d_{Ph-Ph}$) distances between the ligands. **b**, X-rays diffraction (XRD) patterns and Rietveld plots of Hf-DPA:TP-1.6% powder (activated) and Hf-DPA:TP-2.3% films before and after activation. **c**, $^{13}$C CP MAS solid-state NMR of activated Hf-DPA:TP-1.6% powder (up) and the Hf-DPA:TP-2.3% film (down). **d**, Scanning electron microscopy images of activated Hf-DPA:TP-2.3% film deposited on glass substrate at different magnifications. **e**, Profilometry of the thickness of the Hf-TP (pink) and Hf-DPA:TP-2.3% (blue) films from side-to-side. **f-g**, Transmission Electron Microscopy (TEM) images of activated Hf-DPA:TP-1.6% powder at different magnifications. The (11-1) and (200) families of planes are highlighted. **h-i**, TEM image of activated Hf-DPA:TP-2.3% film at different magnifications.

The MOFs exhibit open porosity with high Brunauer-Emmett-Teller (BET) and Langmuir surface areas of 3263 m$^2$/g and 3599 m$^2$/g for Hf-TP and 3213 m$^2$/g and 3541 m$^2$/g for Hf- DPA:TP-1.6%, and pore volumes of 1.22 cm$^3$/g and 1.21 cm$^3$/g, respectively, consistent with the crystal structures (Figs. S14-S17, Supplementary Tables 6-8). Transmission Electron Microscopy (TEM) image of Hf-DPA:TP-1.6% crystalline powder displays well-defined crystals with sharp edges of length ~ 6 μm (Fig. 2f). High magnification images show lattice fringes with interplanar spaces of 17.1 Å and 15.4 Å, in good agreement with the (11-1) and (200) families of planes (Fig. 2g, S18).[45] The good crystallinity of the sample was confirmed by the image of lattice fringes across all the crystals (Fig. S18).

Towards real applications, we fabricated MOF films on a glass substrate using a home-built deposition chamber under optimized solvothermal conditions (Supporting Scheme S1). Homo-ligand MOF film (denoted Hf-TP film) and hetero-ligand MOF films (denoted Hf-DPA:TP-x% films) with increasing amounts of DPA with respect to TP were grown and activated at 110 °C. The compositions of digested Hf-MOF films were checked by $^1$H NMR (Fig. S19-S23, Supplementary Tables 9-11), yielding DPA molar fractions of 0.9%, 1.2%, and 2.3% for the hetero-ligand MOF films and an amount of the monofunctional ligand of 35-47%, higher than that of the crystalline powders. Considering the number of carboxylate units of both di- and mono-functional ligands, the clusters of the metal nodes can be connected by at least 8 difunctional ligands (TP and DPA) out of 12 (>70% connectivity), preserving a high connectivity of the architecture (Supplementary Table 12). Infrared spectra of the MOF films were consistent with the bulk crystalline powders, confirming the coordination of carboxylate units to Hf ions (Fig. S24). $^{13}$C CP MAS NMR spectra displayed signal broadening in the activated films, indicating some local disorder due to the presence of mono-functional ligands (Fig. 2c, Fig. S25–S27). X-ray diffraction patterns of solvent-filled Hf-MOF films confirm the cubic *Fm*-3*m* crystal structure (Fig. 2b, Figs S28-S33). In the activated films, the diffraction peaks broaden and shift toward higher 2-theta values with respect to those in the powders, indicating a shrinkage of the structure, as demonstrated by Rietveld refinement (Fig. S28, S29, Supplementary Table 13). Consequently, the center-to-center and the phenyl-to-phenyl distances between the ligands shorten to 11.0 – 9.8 Å and 6.9 – 6.1 Å, respectively (Fig. S30). This behavior can be ascribed to the replacement of di-carboxylate ligands with mono-carboxylate modulator moieties, which allow for a slight rearrangement of the framework. The peak broadening was analyzed by Rietveld refinement using the double-Voigt approach, which yields a mean domain size of 9±2 nm and 13±2 nm for Hf-TP film (activated) and Hf-DPA:TP-2.3% film (activated), respectively (Fig. S31, S32), suggesting the formation of crystalline nanodomains. Scanning electron microscopy images of the activated thin films showed well-defined, aggregated octahedral crystals with a mean size of 1.4±0.3 μm forming a continuous layer with an average thickness of ~20±0.3 μm (Fig. 2d,e and Fig. S34-S38). TEM images of Hf-DPA:TP-2.3% MOF film confirm a crystal size of ~ 1 μm (Fig. 2h, S39). These crystals exhibit a mosaic microstructure containing nanodomains and dark boundaries, which can be attributed to areas of mass contrast, according to the literature.[46] Higher magnification images enabled highlighting an ordered structure within nanodomains (Fig. 2i, S40).

The luminescence properties of MOF powders and films have been investigated by means of steady-state and time-resolved photoluminescence and scintillation experiments. Figure 3a reports the photoluminescence spectra of the MOF powders under UV excitation at 250 nm. The crystalline powders of Hf-TP show the characteristic emission spectrum of the TP dye, with an emission characteristic lifetime of 236 ps (Fig. S41). The Hf-TP:DPA-1.6% sample, thanks to the fast non-radiative energy transfer from TP ligands, shows the typical blue emission of DPA, with an emission characteristic lifetime of 1.8 ns (Fig. S41). No UV luminescence is observed, thus demonstrating a complete TP-to-DPA energy transfer (Fig. S42b).[44,47] Hafnium has been chosen instead of the previously used zirconium[31] to enhance the absorption cross-section of high energy photons, which increases with a power law by following the atomic number Z of the interacting elements. This choice has several positive effects. First, the presence of Hf is crucial to provide the system with a reasonably good interaction probability with γ-rays. The simulated energy spectra reported in Fig. 3b show how hafnium (Z=72) dramatically increases the photoelectric-effect cross-section with 511 keV photons γ-rays employed in ToF-PET with respect to zirconium (Z=40). Here, the pulse height spectra of an ideal MOF bulk monolith (3×3×20 mm, density 0.77 g cm$^{-3}$, 2.3% DPA content, Supplementary Table 14) are compared to the ones of BGO (density 9.60 g cm$^{-3}$) and a standard polymer (density 1.00 g cm$^{-3}$) with the same size, as reference for dense and light scintillators, respectively. Notably, the Hf-based MOF shows a total fraction of photoelectric events of 7%, which is more than 10 times higher than that of the Zr-based MOF, more than 30 times higher than the polymer and only 6 times lower than the high-density BGO crystal used in ToF-PET scanners. This remarkable result suggests that Hf ions partially compensate for the extremely low density of nanoporous MOFs. Second, a more efficient interaction with γ-rays means a more efficient energy deposition, affecting the LY of the scintillating material. As we recently demonstrated, the presence of the dense hafnium oxide clusters activate a local radiosensitization effect that enhances the low recombination yield of diffusing ionized charges, typical of low-density materials,[48-50] thus enhancing the system LY. The data reported in Fig. 3c show a clear increase of more than 300% in the LY of the MOF powders from Zr-based (Zr-DPA:TP-1.4%, ca. 3600 ph/MeV)[44] to Hf-based systems (Hf-DPA:TP-1.6%, ca. 13500 ph/MeV).

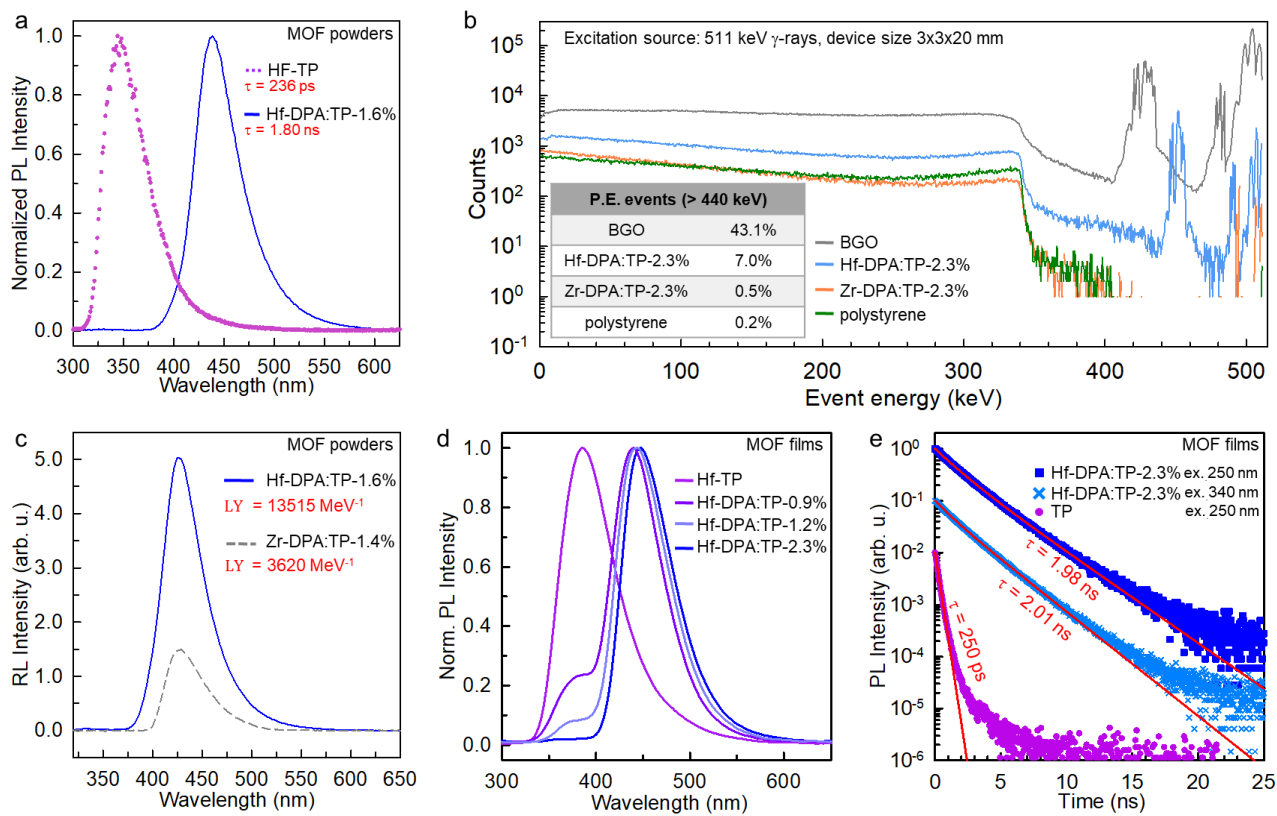

**Figure 3 | a,** Photoluminescence (PL) spectra of MOF powders with (Hf-DPA:TP-1.6%) and without (Hf-TP) doping with the low energy emitting dye DPA under 250 nm excitation. **b,** Energy loss events probability calculated by a Monte Carlo method for 3×3×20 mm simulated monoliths of BGO, Hf-TP:DPA-2.3%, Zr-TP:DPA-2.3% and polystyrene for 511 keV photons. In the inset the fraction of event with an energy deposition larger than 440 keV useful for ToF-PET. **c,** Radioluminescence (RL) spectra of MOF films based on Zr-oxylate (Zr-DPA:TP-1.4%) and Hf-oxylate (Hf-DPA:TP-1.6%) clusters as linking nodes under soft X-rays. **d,** Normalized PL spectra of Hf-DPA:TP-x% films measured as function of the DPA content under 250 nm excitation. **e,** PL intensity decay in time for the Hf-DPA:TP-2.3% and Hf-TP films recorded at the PL peak under different laser excitation wavelengths.

Importantly, the Hf-MOF films successfully retained emissive properties and energy transfer kinetics consistent with those observed in the powder MOF. Figure 3d shows the photoluminescence of the Hf-MOF film series with an increasing DPA amount ranging from 0.0%wt to 2.3%wt, at which concentration the energy transfer yield $\phi_{ET}$ from excited TP ligands to the DPA energy acceptors is ≥ 0.9 (Fig. 3d, S42). Indeed, in MOFs based on conjugated ligands, the hopping-mediated diffusion of molecular excitons (Fig.1) is driven by Förster and Dexter non-radiative energy transfer mechanisms, with a poor influence from phonons.[44] Specifically, in our MOFs, the emissive singlets move fast enough to experience energy transfer before radiative recombination, owing to their high diffusivity $D_S^{TP} = 1.88 \times 10^{-2}$ cm$^2$ s$^{-1}$ (lower limit value considering only on Förster interaction, Supplementary Information, sec. 4) within the framework facilitated by the high density of TP ligands. This allows the calculation of the energy transfer rate $k_{ET}$ to DPA ligands in the rapid diffusion limit by

$$k_{ET} = \frac{4\pi C_{DPA} R_{FS}^6 k_{TP}}{3a^3}, \qquad \text{Eq.1}$$

where $C_{DPA}$ is the DPA ligand concentration expressed in [cm$^{-3}$], $R_{FS}$ is the TP/DPA pair Förster radius (Supplementary Information, sec. 4), $a$ is the minimum center-to-center distance between two molecular excitons in the frameworks and $k_{TP}$ is the decay rate of the TP fluorescence in the absence of DPA (Fig. S41).[51,52] The predicted energy transfer rate and yield $\phi_{ET} = k_{ET}/(k_{ET} + k_{TP})$ perfectly reproduce the experimental values measured as function of the DPA concentration (Fig. S42), thus suggesting that TP-to-DPA energy transfer occurs in the tens-to hundreds of picosecond time regime with rates >> 1 GHz. Figure 3e shows the photoluminescence decay kinetics of Hf-DPA:TP-2.3% film recorded at 430 nm, i.e. at the peak of DPA emission, upon excitation at 340 nm in the DPA absorption band or at 250 nm in the TP absorption band. From both measurements a characteristic photoluminescence intensity decay time of ~ 2 ns was determined. No differences can be noted using different excitation energies, thus demonstrating that the activation of the large Stokes shift blue emission is unaffected by the TP-to-DPA energy transfer. The DPA emission intensity decays with a predominantly single exponential kinetics and a characteristic lifetime similar to the powder (Fig. 3a, S41). The same behavior is observed for the UV emitting homo-ligand Hf-TP film but with a faster characteristic lifetime of 250 ps (Fig.3e).

Figure 4a shows the radioluminescence spectrum of the Hf-TP and Hf-DPA:TP-2.3% films of 22 μm thickness (Fig. 2e) grown on a glass substrate (Fig. S44) under soft X-ray excitation with an average energy of 7 keV. The films show the expected blue and UV luminescence matching the DPA and TP emission spectra, respectively, with a corresponding light yield of 11951 ph MeV$^{-1}$ and 12758 ph/MeV. The associated estimated uncertainty is ±10%. These values have been accurately measured with relative methods and directly measuring the energy deposited in the films by X-rays (Supplementary Information, sec. 6), in order to point out the intrinsic efficiency of the energy-to-photons conversion of the films, i.e. their LY. These results demonstrate that the energy transfer exploited to achieve the Stokes shift does not affect the scintillation properties of the film. Indeed, time-resolved scintillation experiments performed on MOF films as a function of the DPA concentration confirm the rapid diffusion kinetics of the energy transfer during scintillation (Fig. S42). Figure 4b shows the intensity time decay kinetics of the TP scintillation at 340 nm recorded under pulsed X-rays at 14.5 keV as a function of the DPA loading. The drastic acceleration of the emission recombination kinetics observed in the absence of DPA (Fig. 3e) mirrors the faster energy transfer at higher DPA loading levels and allows the calculation of the experimental transfer rate. Notably, for the most performing Hf-DPA:TP-2.3% film the data indicate an energy transfer rate $k_{ET}$ > 50 GHz (< 20 ps), as expected much faster than the approximated estimated value and below the instrumental resolution. The scintillation pulses recorded in the Hf-MOF films under pulsed X-rays show, in both cases, an excellent time response. The Hf-DPA:TP-2.3% scintillation (Fig. 4c) shows a rise time $\tau_{rise}$ = 36 ps and an average decay time $\tau$ = 760 ps, much faster than the photoluminescence lifetime. The Hf-TP film pulse (Fig. 4d) is quicker, with a faster rise time of 28 ps and a faster decay time of 150 ps, again significantly shorter than the corresponding photoluminescence (Fig. 3e). Importantly, ultrafast kinetics and high LY are achived at room temperature, surpassing any organic/polymeric scintillator and most of the recently proposed hybrid systems such as perovskites.[53-57]

The fast kinetics of the rise time is compatible with the exciton diffusivity within the ligands in Hf-TP whilst in the hetero-ligand MOF it is slightly slower due to the picosecond-fast energy transfer step that activates DPA ligands. Moreover, the ultra-fast scintillation decay under pulse X-ray excitation hints the presence of competitive biomolecular processes occurring in the MOF. If the diffusivity of singlets and their density are high enough, there could be a non-zero probability that two fluorescent singlets encounter and collide during their lifetime. The inelastic singlet-singlet annihilation (SSA) can be therefore a competitive channel to the singlets radiative recombination,[58] thus pushing to a faster recombination kinetics owing to a partial quenching of the excited singlet population. Figure 4e shows the rate $k_{SSA}$ for hetero-molecular SSA (dashed line) in the Hf-DPA:TP-2.3% film calculated in the rapid diffusion limit by

$$k_{SSA} = 4\pi(D_S^{TP} + D_S^{DPA})R_{SSA}[S_{DPA}^*], \qquad \text{Eq.2}$$

as function of the energy of the incident photon from 1 to $10^3$ keV (Supplementary Information, sec. 5). Specifically, $[S_{DPA}^*]$ is the density of DPA singlets that can annihilate with the remaining diffusion TP singlets after energy transfer, a process possible since the DPA lifetime is much longer than the TP one. The $[S_{DPA}^*]$ value is estimated as a function of the excitation photon energy by considering a full energy deposition by photoelectric event. $R_{SSA}$ is the singlet-singlet annihilation distance, taken as large as a typical 2 nm to represent the collision of the two exciton molecular electronic orbitals.[59,60] Considering the energy difference between DPA emission and TP absorption that avoids back energy transfer, the diffusivity $D_S^{DPA}$ is considered equal to zero. Notably, the observed scintillation lifetime of 760 ps (red square) is in very good agreement with the predicted value of $\bar{\tau}_{scint}^{DPA} = (k_{scint}^{DPA})^{-1} = (k_{pl}^{DPA} + k_{SSA})^{-1}$ where $k_{SSA}$ is the SSA rate calculated considering the X-ray energy of 14.5 keV employed in the experiment (solid line). This modeling perfectly reproduces also the scintillation kinetics of the homo-ligand MOF. In this case the SSA rate is calculated as

$$k_{SSA} = 8\pi D_S^{TP} R_{SSA}[S_{TP}^*], \qquad \text{Eq.3}$$

where $[S_{TP}^*]$ is now the density of generated TP singlets. Again, with photons at 14.5 keV the predicted value of $k_{SSA}$ allows to calculate an expected scintillation lifetime $\bar{\tau}_{scint}^{TP}$ that matches the experimental value (Fig. 4f). This picture is supported by two evidences. First, in agreement with the much less probable and effective photoelectric interactions in presence of Zr (Fig. 3b), the density of generated singlets in Zr-MOFs is significantly lower than that in Hf-MOFs. Consequently, the SSA in Zr-MOFs is negligible and the scintillation lifetime matches that of the photoluminescence.[44] Second, the observed $k_{SSA}$ rate in Hf-MOFs shows the expected dependency on the energy of the excitation photons that changes the density of singlets. Indeed, at higher energies we observe a faster recombination kinetics, due to the more efficient SSA when a larger number of singlets is generated that quenches more efficiently the MOF emission, in agreement with Eq. 2 (Fig. S45).

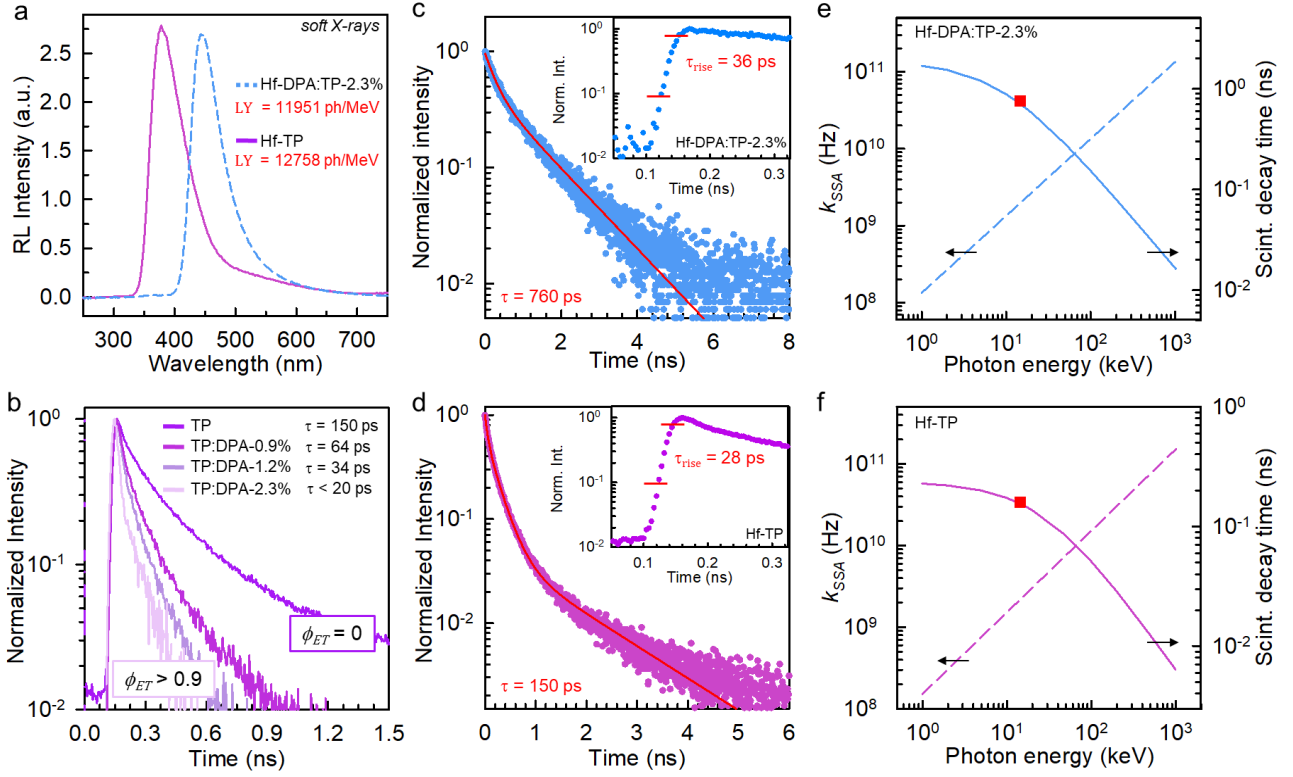

**Figure 4** | **a,** Radioluminescence (RL) spectra of Hf-TP and Hf-DPA:TP-2.3% films under soft X-rays. **b**, Scintillation pulses under soft X-rays recorded at 340 nm in hetero-ligand MOF films as function of the DPA amount. **c, d**. Scintillation pulses for Hf-DPA:TP-2.3% (c) and Hf-TP (d) MOF films recorded at 340 nm and 430 nm, respectively. **e, f**. Theoretical singlet-singlet annihilation (SSA) rate (dashed line) and scintillation lifetime (solid line) as function of the incident photon energy. The red squares mark the experimental lifetime values reported derived from the data in panel c and d.

To evaluate the potential of a new scintillator for fast timing application, we can estimate the ~~potential~~ best temporal resolution achievable. Specifically, in the case of ToF-PET, a crucial figure of merit to be considered is the scanner coincidence time resolution (CTR), i.e. the minimum uncertainty on the detection time difference for the back γ-rays generated by the cancer radiotracer injected in the patient body. The intrinsic, all-optical $CTR^0$ value of a scintillator can be calculated by

$$CTR^0 = 3.33\sqrt{\frac{\tau_{rise}\,\tau}{N}} = 3.33\sqrt{\frac{\tau_{rise}\,\tau_{eff}}{\beta \chi E * LY}}, \qquad \text{Eq. 4}$$

where $N$ is the total number of emitted photons, $\tau_{eff}$ is the effective scintillation decay time, $\beta$ is the light outcoupling efficiency of the scintillator to the photodetector, $E = 511$ keV is the deposited energy by the γ-rays used in ToF-PET, and $\chi$ is the photodetector quantum efficiency at the scintillator emission wavelength.[15,61,62] Considering typical values of 0.3 and 0.25 for $\beta$ and $\chi$, respectively, we calculate a $CTR^0$ as low as 10 ps and 26 ps for Hf-TP and Hf-DPA:TP-2.3% films, respectively (Supplementary Information, sec. 7). In a model ToF-PET detection pixel with size 3×3×20 mm where photoelectric event distribution in space and the travelling path length distribution of the scintillation photons must be considered. These $CTR^0$ values translate into an estimated more realistic CTR of 30 ps and 50 ps for Hf-TP and Hf-DPA:TP-2.3% films, respectively, thus approaching the highly desired and extremely challenging 10 ps value for real-life devices.[4,63] Consequently, these results suggest that the ToF-PET time response using ultrafast scintillating MOFs will be no more limited by the scintillation speed and yield, but just by the limited efficiency of photodetectors and light extraction.

**Conclusion**

In summary, by engineering the composition of Hf-based MOFs we were able to realize solid-state, ultrafast scintillating MOF films with emission both in the UV and the visible range, this latter thanks to a diffusion mediated energy transfer mechanism. The inclusion of high Z hafnium ions in the MOF linking nodes improves the material interaction cross-section with high energy photons partially compensating the low density of the framework, and significantly enhances their scintillation efficiency. Importantly, the crystalline framework, wherein molecular excitons move fast after the recombination of diffusing charges during scintillation, allows the occurrence of bimolecular quenching processes that set the kinetics to the shortest scintillation time achieved so far, down to 150 ps, for MOF based systems. Moreover, the gain in scintillation yield by the inclusion of Hf ions is high enough to keep the light yield of the films around 12000 ph/MeV, making them the best performing fast emitting MOF-based scintillators at room temperature, better than most recent benchmark fast scintillators and commercial systems (Supporting Table 20 and Fig. S46) and thus excellent candidates to develop forefront fast ToF-PET detectors. The efficiency values, the ultrafast emission kinetics and the huge versatility of the MOF composition, that enables to design application-tailored systems and, potentially, control energy diffusion-mediated bimolecular process rates as well as their fabrication in the form of films, make MOF technologically-appealing candidates for the realization of next generation of ultrafast and high performance scintillation counters.

**Methods**

*Materials.* Ligands 2′,5′-dimethyl-[1,1′:4′,1″-Terphenyl]-4,4″-dicarboxylic acid (TP) and 9,10-bis(4-carboxyphenyl)anthracene (DPA) were purchased from Fluorochem. Hafnium chloride ($Cl_4Hf$), benzoic acid (BA, 99.5%), hydrochloric acid (HCl), tetrahydrofuran (THF), dry dimethylformamide (DMF) and chloroform ($CHCl_3$) were purchased from Merck. Deuterated dimethyl sulfoxide (DMSO_d6) and deuterated trifluoroacetic acid (TFA_d) were purchased from Merck.

*Synthesis of Hf-TP and Hf-DPA:TP-x% powder.* MOF nanocrystals were synthesized under solvothermal conditions modulated by benzoic acid. $HfCl_4$ was dispersed with 2',5'-dimethyl-(1,1':4',1"-terphenil)-4,4"-dicarboxylic acid, benzoic acid and x% of 9,10-bis(4-carboxyphenyl) anthracene in DMF. The mixture was heated at 120°C for 20h and the whitish powder obtained was collected and washed with fresh DMF and exchange with $CHCl_3$. The solvent was removed and the sample activated at 110°C for 18 h under high vacuum.

*Synthesis of Hf-TP films.* Homo-ligand MOF films were synthesized under solvothermal conditions modulated by benzoic acid. $HfCl_4$ was dispersed in DMF and sonicated for 15 minutes. The solution of $HfCl_2$ and DMF was added into the reactor containing 2',5'-dimethyl-(1,1':4',1"-terphenil)-4,4"-dicarboxylic acid and benzoic acid and additional DMF. The reactor was sonicated for 5 minutes and the sample holder with a glass slip was inserted. The system was tightly sealed and heated at 120°C for 24h. The MOF-substrate was extracted, washed with fresh DMF and activated at 110°C for 18h under high vacuum.

*Synthesis of Hf-DPA:TP-x% films.* Hetero-ligand MOF films were synthesized under solvothermal conditions modulated by benzoic acid. 9,10-bis(4-carboxyphenyl) anthracene was dispersed in DMF and heated up to 120°C for the complete dissolution of the ligand. Separately $HfCl_4$ was dispersed in DMF and sonicated for 15 minutes. The solution of $HfCl_2$ and DMF was added into the reactor containing 2',5'-dimethyl-(1,1':4',1"-terphenil)-4,4"-dicarboxylic acid and benzoic acid and, lastly, the hot solution of DPA in DMF was poured. The reactor was sonicated for 5 minutes and the sample holder with a glass slip was inserted. The system is tightly sealed and heated at 120°C for 24h. The MOF-substrate was extracted, washed with fresh DMF and activated at 110°C for 18h under high vacuum.

*Powder X-ray diffraction (PXRD).* PXRD patterns were performed before and after the activation of the samples in a Rigaku SmartLab powder diffractometer using Cu-Kα radiation, 40 kV, 30 mA over a range for 2θ of 2.0 - 70.0° with a step size of 0.02° and a scan speed of 1.0°·min$^{-1}$ equipped with a vacuum chamber (Anton Paar TTK 600) operating with liquid nitrogen cooling system. PXRD patterns under vacuum were collected under dynamic vacuum (p < 1 x 10-2 mmHg) between 293 K and 383 K. PXRD patterns were also performed using a Miniflex diffractometer with Cu-Kα radiation, 40 kV, 20 mA over a range for 2θ of 2.0 - 60.0° with a step size of 0.02° and a scan speed of 1.0°·min$^{-1}$.

*Rietveld Refinement.* Rietveld structural refinements of the X-ray data were performed using the TOPAS-Academic64 V6 software package. The initial model was constructed from a previously solved hetero-ligand Zr-DPA:TP-x% MOF model by replacing the Zr with Hf. The rigid bodies were set up such that the oxide and hydroxide moieties could adjust their relative position on the metal node. We also accounted for reorientation of the carboxylates, phenyl ring, dimethyl phenyl ring and the anthracene moieties. The occupancies for the composition of the ligand site were set to account for benzoate, modulator (BA), TP and DPA. $^1H$ solution NMR determined the occupancies of the different species. The structure was modelled as a disordered system with a space group *Fm*-3*m* for the final Rietveld refinement. The background was fitted and refined using a Chebyshev polynomial with 20 coefficients in the PXRD trace range from 2° to 60° 2theta with baseline shift refinement. The "Simple_Axial_Model" accounted for the asymmetry in the peaks, especially at low 2-theta values. The peaks were fitted using a PearsonVII " PVII " function. For the XRD of the "layer-grown" MOF, a three-phase refinement was used with the unit cells of each phase allowed to be refined. Additional corrections include March-Dollase preferred orientation corrections on the (1 1 1), (2 0 0) and (3 1 1) reflections.

*Solution NMR Analysis.* NMR spectra were obtained using a Bruker Avance Neo spectrometer operating at 9.4 T (400 MHz) at 25 °C. $^1$H NMR chemical shifts (δ) were reported in ppm relative to the proton resonances resulting from incomplete deuteration of the NMR solvents. According to a literature method, the composition of the hetero-ligand MOFs was quantified from the analysis of $^1$H NMR spectra.

*Thermogravimetric analysis.* Thermal Gravimetric Analysis (TGA) was performed using a Mettler Toledo Star System 1 equipped with a gas controller GC10. Samples were outgassed overnight at 100°C under high vacuum ($p \leq 3*10^{-6}$ bar) to remove adsorbed species. The experiments were conducted by applying a thermal ramp from 30°C to 1000°C and a scan rate of 10°C/min in dry air

*FT-IR spectroscopy.* Fourier-transform infrared spectroscopy was performed using a Thermo Scientific Nicolet iS20 spectrometer equipped with a Smart iTX Accessory for ATR measurements. The spectra were collected using 128 scans with a resolution of 4 cm$^{-1}$ between 525 cm-1 and 4000 cm-1 on activated samples in air.

*Gas sorption analysis.* Before sorption analysis, the samples were activated at 100°C overnight to remove solvent from the pores under high vacuum ($p \leq 3*10^{-6}$ bar). Adsorption experiments were performed using N$_2$ with a purity of 5.0 and the adsorption isotherms at 77 K were collected on a Micromeritics 3Flex Adsorption Analyzer equipped with micropore analysis ports. Low-temperature analysis at 77 K and 195 K were conducted using liquid nitrogen bath. The N$_2$ adsorption isotherms at 77 K were analysed using Flex software. BET and Langmuir surface area were calculated between 0.015 and 0.065 p/p°. The pore size distributions were calculated using the NLDFT theory and the HS-2D carbon slit pore model.

*High-Resolution SEM.* Scanning electron microscopy (SEM) images were collected using a Zeiss Gemini 500 microscope, operating at 5 kV and with a Thermo Fisher Phenom G6 SEM, operating at 10 kV. The microcrystalline powders were deposited on a conductive tape, dried under a high vacuum, and sputtered with gold before the analysis (10 nm, nominal thickness). Images were also collected with a Thermo Fisher Phenom G6 SEM, operating at 10 kV

*Profilometry.* The thickness of the samples was measured using a Dektak 8 Stylus Profiler (Veeco Instruments, Inc.) equipped with an N-Lite low force sensor with a computerized surface and an accuracy in the range of 0.20 microns. The samples were characterized using a 2.5-µm-radius tip with a recording speed of 0.150 mm/s and a loading force of 0.03 mg.

*Solid state NMR.* $^{13}$C and $^1$H solid-state NMR experiments were carried out at 75.5 and 300.1 MHz, respectively, with a Bruker Avance Neo instrument operating at a static field of 7.04 T equipped with a 4 mm double resonance MAS probe. $^{13}$C ramped-amplitude Cross Polarization (CP) experiments were performed at room temperature at a spinning speed of 12.5 kHz using a recycle delay of 5 s and a contact time of 2 ms. The 90° pulse for the proton was 2.5 µs. Crystalline polyethylene was taken as an external reference at 32.8 ppm from TMS. Quantitative $^1$H SPE MAS NMR spectra were performed at room temperature at a spinning speed of 12.5 kHz using a recycle delay of 20 s. The 90° pulse for the proton was 2.4 µs. The $^1$H chemical shift was referenced with respect to adamantane.

*Photoluminescence studies.* Time scale time-resolved photoluminescence experiments had been performed by using as excitation source a pulsed laser LED at 340 nm (3.65 eV, EP-LED 340 Edinburgh Instruments, pulse width 120 ps) and a pulsed laser LED at 250 nm (4.95 eV, EP-LED 250 Edinburgh Instruments, pulse width 77 ps) coupled to a Horiba Horiba-Jobin Yvon Simphony monochromator and, as photodetector, to and a hybrid photomultiplier tube PMA Hybrid 07 (Picoquant GmbH) coupled to a 4 ps resolution Pico Harp 300 TCSPC module. A custom-made 5000M spectrofluorimeter (Horiba-Jobin Yvon) equipped with a TBX-04 photon counting detector (IBH Scotland) and a single grating monochromator was used for the photoluminescence spectra measurements within the 200–800 nm range.

*Radioluminescence studies.* The samples were excited by unfiltered X-ray irradiation using a Philips PW2274 X-ray tube, with a tungsten target, equipped with a beryllium window and operated at 20 kV. At this operating voltage, X-rays are produced by the Bremsstrahlung mechanism, superimposed to the L and M transition lines of tungsten due to the impact of electrons generated through a thermionic effect and accelerated onto the tungsten target. Cryogenic radioluminescence measurements are performed in the 10−370 K interval. Radioluminescence of powders has been recorded on powder samples of 1 mm thickness in an aluminum sample holder.

*Scintillation studies.* The scintillation light pulses had been recorded in time correlated single photon counting (TCSPC) mode under pulsed X-ray excitation. For this purpose, an X-ray Tube (XRT) N5084 of Hamamatsu was used, activated by a pulsed 405 nm laser (pulse width EP-LED 250 Edinburgh Instruments, pulse width 120 ps). The X-rays energy spectrum was a bremsstrahlung continuous spectrum extending up to 40 keV (as the operating voltage is 40 kV, pulse width 80 ps) with an additional pronounced peak ≈ 9 keV due to Tungsten L-characteristic X-ray photons. As photodetector, a hybrid photomultiplier tube PMA Hybrid 07 (Picoquant GmbH) coupled to a 4 ps resolution Pico Harp 300 TCSPC module was used.

*Light yield measurements under soft X-rays.* The MOF LY was measured by relative methods using the EJ276D polymeric scintillators as reference (LY 8600 ph MeV$^{-1}$, experimentally confirmed in a 1 mm thick sample by comparison with a 1 mm BGO crystal, LY 10000 MeV$^{-1}$). For powders, 1 mm thick samples have been employed in an aluminum sample holder. The EJ-276D has been grinded to powders in order to have a quantitative and accurate comparison. For films, the EJ-276D has been sliced down to a 100 μm thickness and 20×20 mm shape to match the MOF/glass heterostructure and finely scratched on the surface to mimic the MOF layer scattering. The LY has been evaluated by comparing the relative radioluminescence integrated intensity recorded with the same experimental conditions. Given the low thickness of the film samples, the radioluminescence intensities have been corrected by the effective absorption of the excitation X-rays beam. To do that, we measure the integrated radioluminescence emission of a 5 mm BGO crystal (LY=10000 ph/MeV) under the excitation beam (100% absorption at 7 keV) and we compare its emission intensity when the beam goes through our film samples before hitting the BGO. The relative decrease of the BGO emission intensity allows us to estimate the relative absorptance of the excitation beam by any film, thus enabling a proper correction of the recorded radioluminescence intensity and therefore a proper evaluation of the films (Supplementary Information, sec. 6)

*Radiation/Matter Monte Carlo Simulation.* In order to evaluate the stopping power of MOFs for 511 keV γ-rays, a Monte Carlo simulation of the multilayer and bulk scintillators had been performed by means of the FLUKA code.[64] The simulated scintillators have a size of 3×3×20 mm, and their composition in terms of atomic weights and density had been fully reproduced. The 511 keV photons isotropic source had been put at a distance of 0.5 cm from the scintillator surface. The simulation output had been analyzed on an event-by-event basis through dedicated user-routines developed on purpose.

## Data availability

The data that support the plots within this paper and other findings of this study are available from the corresponding author upon reasonable request. No custom code has been developed for computational modeling and simulations.

## Acknowledgements

We acknowledge financial support from the Italian Ministry of University (MUR) through grant no. PRIN 2020-SHERPA no. H45F2100343000, grant and grant MINERVA – LuMIminesceNt scintillating hEterostructures foR advanced medical imaging no. H25E22000490006nd, PRIN-2022 HYSTAR no. H53D23004720006 and from Lombardy Region through the 'Enhancing photosynthesis' award, no. H45F21002830007. We thank Dr. Chiara L. Boldrini at the Mibsolar center of the University of Milano-Bicocca for the help with profilometry experiments. We thank Dr. Francesca Cova at the Department of Materials Science University of Milano-Bicocca of the for the help in setting up X-rays absorption measurements.

## Authors contribution